\documentclass[12pt]{article}
\usepackage{color} 
\usepackage{epsf}
\usepackage{amsmath,amssymb,graphicx}
\usepackage{bm}

\begin{document}

\renewcommand{\theequation}{\thesection.\arabic{equation}}

\begin{titlepage}

\title{Large Scale Quasi-geostrophic Magnetohydrodynamics}
\date{}
\author{Alexander M. Balk\\
Department of Mathematics\\ University of Utah \\
Salt Lake City, UT 84112}
\maketitle

\begin{abstract}

We consider the ideal magnetohydrodynamics (MHD) of a shallow fluid
layer on a rapidly rotating planet or star.
The presence of a background toroidal magnetic field is assumed, 
and the ``shallow water" beta-plane approximation is used. 
We derive a single equation for the slow large length scale dynamics. 
The range of validity of this equation fits the MHD of the
lighter fluid at the top of Earth's outer core.  
The form of this equation is similar 
to the quasi-geostrophic (Q-G) equation
(for usual ocean or atmosphere),
but the parameters are essentially different. 
Our equation also implies the inverse cascade; 
but contrary to the usual  Q-G situation, 
the energy cascades to smaller length scales, 
while the enstrophy cascades to the larger scales. 
We find the Kolmogorov-type spectrum for the inverse cascade. The
spectrum indicates the energy accumulation in larger scales.
In addition to the energy and enstrophy, 
the obtained equation possesses an extra invariant. Its presence
is shown to imply energy accumulation in the $30^\circ$ sector around
zonal flow. With some special energy input, the extra invariant
can lead to the accumulation of energy in zonal flow; 
this happens if the input of the extra invariant is small, 
while the energy input is considerable.

\end{abstract}


\thispagestyle{empty}
\end{titlepage}
 
\section{Introduction}
\setcounter{equation}{0}
\label{Sect: Intro}

The behavior of various stars and planets crucially depends on MHD of
some shallow fluid layer (fluid shell); 
examples are the Solar {\it tachocline} (e.g. \cite{Tachocline})
or the Earth {\it ocean of the core} 
(the layer of a lighter fluid at the top of
the outer core, e.g.\ \cite{Brag07}). 
Such situations
can be studied using the system of ``Shallow Water'' MHD
\cite{gilman}. 

The system includes five (scalar) evolution equations
\begin{subequations}\label{SMHD}
\begin{eqnarray}
{\bf V}_t+({\bf V}\cdot\nabla){\bf V}+{\bf f}\times{\bf V}&=&
      -g\nabla H+({\bf B}\cdot\nabla){\bf B},\quad\quad\label{Vt}\\
H_t+\nabla\cdot(H {\bf V})&=&0,\label{Ht}\\
{\bf B}_t+({\bf V}\cdot\nabla){\bf B}&=&({\bf B}\cdot\nabla){\bf V}, \label{Bt}
\end{eqnarray}
subject to the constraint
\begin{eqnarray}
\nabla\cdot(H {\bf B})&=&0. \label{cnstrain}
\end{eqnarray}
\end{subequations}
The equations (\ref{SMHD}) are written in the Cartesian geometry 
of the $(x,y)$-plane tangent to the fluid shell.
The fluid velocity ${\bf V}$, the magnetic field ${\bf B}$, 
and the fluid depth $H$
are unknown functions of $x,y$, and time $t$.
The magnetic field is normalized to have velocity units.
In (\ref{SMHD}), the vectors ${\bf V}$ and ${\bf B}$ have
only two non-zero components, in the plane $(x,y)$;
the momentum equation (\ref{SMHD}a) includes the Coriolis force
with {\bf f}=[0,0,f(y)]; $g$ is the gravity constant.
Throughout the paper, subscripts $x,y,t$ denote partial derivatives, 
while superscripts $x,y$ denote vector components.
The system (\ref{SMHD}) presents significant simplification over the
full system of three-dimensional MHD; and at present, the system
(\ref{SMHD}) is well established; see e.g.
\cite{Schecter01,sterk01,Dellar03,
ZaqOliBalShe,HengSpitk09,Umurhan13,Petrosyan13,ztln13}.

The system (\ref{SMHD}) has steady solution 
\begin{equation}\label{bkgrd}
	{\bf V}=0,\; H=H_0,\; B^x=B_0,\; B^y=0;
\end{equation}
it describes a resting fluid layer of uniform thickness $H_0$,  
penetrated by the uniform toroidal magnetic field $B_0$. 
In the $f$-plane approximation [when $f(y)\equiv f_0$ constant],
the dynamics (\ref{SMHD}) on the background (\ref{bkgrd}) 
has waves of two types  \cite{Schecter01}:
\begin{enumerate}
\item fast waves --- ``magnetogravity'' branch; 
they reduce to the gravity (or Poincar\'{e}) waves 
(familiar in Geophysical fluid dynamics, e.g. \cite{Va}) if $B_0\rightarrow 0$;
\item slow waves --- ``Alfven'' branch; 
their frequencies vanish if $B_0\rightarrow 0$.
\end{enumerate}
When the latitudinal variation of the Coriolis parameter is taken into account, 
i.e. on the beta-plane, $f(y)=f_0+\beta y$, the slow ``Alfven'' branch 
further splits into two sub-branches \cite{ZaqOliBalShe}; 
if the Alfven speed $B_0$ is much smaller than 
the gravity wave speed $c_g=\sqrt{g H_0}$ 
and if the typical length scale ${\mathcal L}$ is sufficiently large:
\begin{equation}\label{ell}
{\mathcal L}\gg \ell\equiv \sqrt{{B_0}/{\beta}},
\end{equation}
then one sub-branch represents Rossby waves, slightly modified by the
magnetic field, and the other sub-branch represents much slower waves.

In the present paper, 
we show that in a certain regime (Sect.\ \ref{Sect:regime}),
the slowest mode (the slow sub-branch of the slow ``Alfven'' branch) 
obeys a closed nonlinear equation [Eq (\ref{Eq}) below].
The form of this equation is similar
   to the usual Q-G equation (from Geophysical fluid dynamics) 
or to the Hasegawa-Mima equation (from Plasma physics), 
but the parameters are very different (Sect. \ref{Sect: Eq}).
The obtained equation describes the dynamics 
of interacting waves with dispersion law
\begin{eqnarray}\label{Omega}
\Omega_{\bf k}=\frac{B_0^2}{\beta}\frac{p\;k^2}{1+\rho^2 k^2}\,,
\quad\mbox{ where }\quad \rho\equiv\frac{B_0 f_0}{\beta c_g}
\end{eqnarray}
and ${\bf k}=(p,q)$ is the wave vector ($k^2=p^2+q^2$).
In its form, the function $\Omega_{\bf k}$ differs from 
the usual Rossby wave dispersion law only by a Doppler shift.  
The nonlinearity in the obtained equation 
is the same as in the usual Q-G one.

We see in Sect. \ref{Sect: Estim} that the regime of validity 
for this equation is realized in the Earth case.

Unlike the system (\ref{SMHD}), 
the obtained equation, in addition to the energy, 
conserves another positive-definite quadratic quantity --- enstrophy.
This leads to the inverse cascade. 
However, contrary to the usual Q-G situation, 
the energy cascades to smaller scales, 
while the enstrophy cascades to larger scales --- Sect. \ref{Sect:Cascds}.

Concentrating on the inverse cascade, we use dimensional considerations
to find the Kolmogorov-type spectrum
\begin{eqnarray}\label{Ek}
E_k\sim \;(\beta\,Q)^{1/2}\;k^{-3/2},
\end{eqnarray}
where $Q$ is the enstrophy flux.
The energy integral has infrared divergence on this spectrum, 
indicating \cite{ZakhKS} that the energy accumulates at larger scales
(similar to the sea wave turbulence) ---  Sect. \ref{Sect:Kolmgrv}.

We further show that the obtained equation has 
the extra, adiabatic-like, invariant (Sect. \ref{Sect:Inv})
\begin{eqnarray}\label{TheIntegral}
I = \int \frac{\eta({\bf k})}{\Omega_{\bf k}}\; E_{\bf k}\; d{\bf k},
\end{eqnarray}
where $E_{\bf k}$ is the energy spectrum, and
\begin{eqnarray}\label{eta} 
\eta({\bf k})\equiv\arctan\frac{q+p\sqrt{3}}{\rho\, k^2}
                  -\arctan\frac{q-p\sqrt{3}}{\rho\, k^2}\,.
\end{eqnarray}
The presence of this invariant implies 
the concentration of large-scale energy
in the $30^\circ$ sector around zonal flow 
(Sect. \ref{Sect:InvrsCascd})
and can even lead to the formation of zonal flow
(Sect. \ref{Sect:Zonl}).

\section{Equation for Large Scale Q-G MHD}
\setcounter{equation}{0}
\label{Sect: Eq}

Studying dynamics on the background (\ref{bkgrd}), we assume
\begin{equation*}
{\bf V}=(v^x,v^y),\;\;H=H_0+h,\;\;{\bf B}=(B_0+b^x,b^y).
\end{equation*}

\subsection{Preliminary Considerations: Dominate Balances}
\label{Sect: DominateBal}

We assume the following three dominate balances 
in the equations (\ref{SMHD}abc).

\paragraph{} {\it Geostrophic balance:} In the momentum equation (\ref{Vt}),  
the Coriolis force and the pressure gradient 
(due to the fluid height) balance each other, 
while dominating all other terms; 
the next biggest term is assumed to be 
the one with the background magnetic field 
\begin{subequations}\label{balances}
\begin{eqnarray}\label{QG}
f v^x\approx -g h_y+ B_0 b^y_x,\quad
-f v^y\approx - g h_x+B_0 b^x_x.\quad
\end{eqnarray}
\paragraph{} In the continuity equation (\ref{Ht}) 
after substitution of the velocity (\ref{QG}), 
{\it the term with gradient $\beta\equiv f'(y)$ balances 
the term with background magnetic field $B_0$}
\begin{eqnarray*}
-\frac{\beta g}{f^2}h_x+\frac{B_0}{f} (b^y_x-b^x_y)_x \approx 0.
\end{eqnarray*}
This equation can be integrated in $x$; 
assuming that $b^y_x-b^x_y$ and $h$ vanish at $x=\infty$, 
we find
\begin{eqnarray}\label{betaBal}
h\approx\frac{f B_0}{g \beta} (b^y_x-b^x_y).
\end{eqnarray}
\paragraph{} {\it Wave balance} in the induction equation (\ref{Bt}):
The term with time derivative balances the term with background
magnetic field 
\begin{eqnarray}\label{waveBal}
b^x_t\approx B_0 v^x_x,\quad
b^y_t\approx B_0 v^y_x. 
\end{eqnarray}
\end{subequations}

The wave balance (\ref{waveBal}) implies
\begin{eqnarray}\label{jzeta}
 j_t\approx B_0\zeta_x\,,
\end{eqnarray}
where $j\equiv b^y_x-b^x_y$ and $\zeta\equiv v^y_x-v^x_y$.
In the equation (\ref{jzeta}), 
we express $j$ in terms of $h$ using (\ref{betaBal})
and substitute geostrophic velocity 
[given by (\ref{QG}) without the magnetic correction] into $\zeta$
\begin{eqnarray*}
h_t\approx \frac{B_0^2}{\beta} \Delta h_x.
\end{eqnarray*}
This equation describes linear waves with dispersion relation
\begin{eqnarray}\label{disp0}
\omega\approx  \frac{B_0^2}{\beta} pk^2 \quad
[h\propto e^{i(p x + q y - \omega t)}].\quad
\end{eqnarray}
Considering corrections 
beyond the dominate balances, 
we will find the interaction of these waves 
and see the energy transfer between waves 
with different wave vectors ${\bf k}$. 
We will also find correction to the dispersion relation (\ref{disp0}), 
which allows for the presence of the extra invariant 
(in addition to the energy and enstrophy).

\bigskip
Let us consider characteristic scales
\begin{eqnarray}\label{scales}
{\partial}/{\partial x},{\partial}/{\partial y} 
\sim {1}/{\mathcal L},\quad
{\partial}/{\partial t} \sim {1}/{\mathcal T},\quad
h \sim {\mathcal H},\quad\nonumber\\
{\bf v}\equiv [v^x,v^y] \sim {\mathcal V},\quad 
\quad
{\bf b}\equiv [b^x,b^y] \sim {\mathcal B}.\quad\quad
\end{eqnarray}
The three balances (\ref{balances}) imply respectively that
\begin{eqnarray}\label{magnitudeBal}
f\,{\mathcal V}\sim \frac{g{\mathcal H}}{{\mathcal L}},\quad\quad
{\mathcal H}\sim \frac{f B_0 {\mathcal B}}{g\beta{\mathcal L}},\quad\quad
\frac{\mathcal B}{\mathcal T}\sim \frac{B_0 {\mathcal V}}{{\mathcal L}}.
\end{eqnarray}

\subsection{Momentum and Induction Equations}

\subsubsection{Quasi-geostrophic velocity}

The two $x,y$-components of the momentum equation (\ref{Vt}) 
can be written in the form
\begin{eqnarray} \label{QGv}
f v^x&=&-g h_y + B_0 b^y_x
  +{\bf b}\cdot\nabla b^y - v^y_t - {\bf v}\cdot\nabla v^y,\quad\nonumber\\
f v^y&=&g h_x - B_0 b^x_x
       -{\bf b}\cdot\nabla b^x + v^x_t + {\bf v}\cdot\nabla v^y;\quad\\
&&\hspace{-1cm}\sim f_0 {\mathcal V}\left\{1+\frac{\beta{\mathcal L}}{f_0}
\left[1\;+\;\frac{\mathcal B}{B_0}\;+\;\frac{\ell^4}{\mathcal L^4}
 \;+\;\frac{\mathcal B}{B_0}\frac{\ell^4}{\mathcal L^4}\right]\right\};\nonumber
\end{eqnarray}
Below these equations we wrote the magnitudes of the
corresponding terms (in the right-hand sides), 
using the characteristic scales (\ref{scales}); 
these magnitudes follow from the dominant balances scaling
(\ref{magnitudeBal}); $f_0\equiv f(0)$.
For instance, to estimate the last terms 
in the right-hand sides of equations (\ref{QGv}),
we note ${\mathcal V}\sim B_0{\mathcal B}/\beta {\mathcal L}^2$ 
and find
\begin{eqnarray*}
\frac{\mathcal V^2}{\mathcal L}\sim
f_0{\mathcal V}\frac{\mathcal V}{f_0{\mathcal L}}\sim
f_0{\mathcal V}\frac{B_0{\mathcal B}}{f_0\beta{\mathcal L^3}}\sim
f_0{\mathcal V}\frac{\beta {\mathcal L}}{f_0}
\frac{\mathcal B}{B_0}\frac{B_0^2}{\beta^2{\mathcal L}^4};
\end{eqnarray*}
now recall the definition of $\ell$ in (\ref{ell}).

\subsubsection{Magnetic potential}
\label{Sect: MagnPot}

The constraint (\ref{cnstrain}) implies the existence of 
a function $A(x,y,t)$ such that
\begin{eqnarray*}
	H B^x = -H_0 A_y,\quad\quad H B^y = H_0 A_x
\end{eqnarray*}
(here, the constant factor $H_0$ is just for normalization).
There is a simple evolution equation for the function $A$
\begin{eqnarray}\label{At}
	A_t+v^x\, A_x+v^y\, A_y=0;
\end{eqnarray}
a similar equation was derived by Gilman \cite{GilmanI67} for stratified
flow; exactly Eq.\ (\ref{At}) is given (without detailed derivation) 
by Zeitlin \cite{ztln13}. 

Let $A_0(y)$ be the part of the potential $A$ that corresponds to the
background (\ref{bkgrd}): $A=A_0+a$, and
\begin{eqnarray*}
(H_0+h)(B_0+b^x)&=&-H_0 (A_0'+a_y),\\
(H_0+h)b^y&=&H_0 a_x.
\end{eqnarray*}
So, $A_0'=-B_0$, and
\begin{align}\label{b}
b^x&=-a_y&-&  &\frac{B_0}{H_0}h&  &+&  
&\frac{(H_0 a_y+B_0 h)h}{H_0(H_0+h)},\nonumber\\
b^y&=a_x &  &  &               &  &-&
&\frac{a_x h}{H_0+h};\\
&\sim{\mathcal B}&
+&  &{\mathcal B}\frac{\beta {\mathcal L}}{f_0} 
\frac{\rho^2}{\mathcal L^2}
&  &+&  &{\mathcal B}\frac{\beta{\mathcal L}}{f_0}
\frac{\mathcal B}{B_0}
\frac{\rho^2}{\mathcal L^2}.\nonumber
\end{align}
Below the two equations, we wrote the magnitudes of the corresponding
terms;
to find these, one can note that the magnitude of the potential $a$ is
${\mathcal A}={\mathcal B}{\mathcal L}$, and 
\begin{eqnarray*}
\frac{\mathcal H}{H_0}\sim\frac{\beta{\mathcal L}}{f_0}
\frac{\mathcal B}{B_0}\frac{\rho^2}{\mathcal L^2}.
\end{eqnarray*}

According to equation (\ref{At}),
\begin{eqnarray}\label{at}
	a_t+v^x\, a_x+v^y\, a_y=B_0 v^y.
\end{eqnarray}

\subsection{Continuity Equation. Approximation}
\label{Sect: ContEq}
The continuity equation (\ref{Ht}) can be written in the form
\begin{eqnarray}\label{contBeta}
	h_t&+&\frac{H_0}{f}[(fv^x)_x+(fv^y)_y-\beta v^y]\nonumber\\
           &+&(hv^x)_x+(hv^y)_y=0,
\end{eqnarray}
which will be used to find $v^y$.

\subsubsection{Small parameters. Considered regime}
\label{Sect:regime}
So far, no approximation has been made.
Now we find the fluid velocity by perturbations 
using the small parameters 
\begin{eqnarray}\label{SmallParam}
\frac{\mathcal B}{B_0},\quad
\frac{\beta{\mathcal L}}{f_0},\quad
\frac{\rho^2}{\mathcal L^2},\quad
\frac{\ell^2}{\mathcal L^2},
\end{eqnarray}
which appear in the asymptotic expansions of the right-hand sides 
in equations (\ref{QGv}) and (\ref{b}).
To guarantee the smallness of parameters (\ref{SmallParam}), 
we assume weak nonlinearity ${\mathcal B}\ll B_0$ and consider regime
\begin{eqnarray}\label{regime}
\ell, \; \rho\; \ll\; {\mathcal L} \;\ll R_0,
\end{eqnarray}
where $R_0$ is the radius of the spherical fluid shell 
[the system (\ref{SMHD}) describes dynamics 
in the plane tangent to this shell]. 
In Earth's case, $R_0$ is the radius of the core-mantle boundary. 
The possibility to use the beta-plane approximation is due 
to the right inequality in (\ref{regime});
it also means that $\beta {\mathcal L}\ll f$ 
(since at mid latitudes, $\beta\sim f/R_0$). 
The left inequality in (\ref{regime})
means that the length scale ${\mathcal L}$ exceeds both lengths
$\ell$ and $\rho$, defined in (\ref{ell}) and (\ref{Omega}).

Let us suppose, for instance, that all four ratios (\ref{SmallParam}) scale
proportional to the same small parameter $\epsilon\rightarrow 0$. 
Then according to (\ref{b}),
\begin{eqnarray}\label{bAsy}
  b^x=-a_y+{\mathcal B}\; O(\epsilon^2),\quad
  b^y=a_x+{\mathcal B}\; O(\epsilon^2)
\end{eqnarray}
and according to (\ref{QGv}),
\begin{subequations}\label{vAsy}
\begin{eqnarray}
f \, v^x&=&-g \, h_y+B_0 \, b^y_x+{\bf b}\cdot\nabla b^y
                         +f_0{\mathcal V}\, O(\epsilon^3),\quad\\
f \, v^y&=&g \, h_x-B_0 \, b^x_x-{\bf b}\cdot\nabla b^x
                         +f_0{\mathcal V}\, O(\epsilon^3).
\end{eqnarray}
\end{subequations}

\subsubsection{Fluid velocity}
Now from the continuity equation (\ref{contBeta})
\begin{eqnarray}\label{hVy}
v^y=\frac{B_0}{\beta}\Delta a_x
+\frac{1}{\beta} \{a,\Delta a\}
+\frac{f}{H_0\beta}h_t+{\mathcal V}\;O(\epsilon^2)\quad\quad
\end{eqnarray}
(we use the Jacobian notation: $\{F,G\}\equiv F_xG_y-F_yG_x$ for
arbitrary functions $F,G$). 
Then from the equation (\ref{vAsy}b)
\begin{eqnarray*}
 \frac{f B_0}{\beta}\Delta a_x=g h_x+f_0{\mathcal V}\,O(\epsilon),
\end{eqnarray*}
which we integrate in $x$ (assuming $\Delta a$ and $h$ 
both vanish at $x=\infty$)
\begin{eqnarray}\label{h}
h=\frac{f B_0}{g \beta}\Delta a+{\mathcal H}\,O(\epsilon).
\end{eqnarray}
We substitute (\ref{h}) into equation (\ref{hVy}) 
\begin{eqnarray}\label{Vy}
v^y=\frac{B_0\Delta a_x}{\beta}
+\frac{\{a,\Delta a\}}{\beta}
+\frac{f^2 B_0\Delta a_t}{g H_0 \beta^2}+{\mathcal V}O(\epsilon^2)\quad\quad
\end{eqnarray}
\bigskip
and into equation (\ref{vAsy}a)
\begin{eqnarray}\label{Vx}
v^x=-\frac{B_0}{\beta}\Delta a_y+{\mathcal V}\;O(\epsilon).
\end{eqnarray}

\subsubsection{The equation for magnetic potential}

Finally, we substitute the fluid velocity (\ref{Vx}), (\ref{Vy}) into
the equation (\ref{at}); and neglect higher order terms
\begin{eqnarray}\label{Eq}
(a-\rho^2 \Delta a)_t=
\frac{B_0^2}{\beta}\Delta a_x
+2\,\ell^2\{a,\Delta a\}.
\end{eqnarray}
Note the factor 2 in (\ref{Eq}): Half of the nonlinear term
comes from the left-hand side of Eq.\ (\ref{at}), and the
other half --- from the expression (\ref{Vy}).

The form of the equation (\ref{Eq}) is very similar 
to the Q-G potential vorticity equation for usual ocean and atmosphere
(e.g. \cite{Va}). Indeed,
the Galilean transformation
\begin{eqnarray*}
a(x,y,t)=\breve{a}(x-u\, t, y, t)
\end{eqnarray*}
turns (\ref{Eq}) into
\begin{eqnarray}\label{modifiedEq}
(\breve{a}-\rho^2 \Delta \breve{a})_t=u\, \breve{a}_x +
\left(\frac{B_0^2}{\beta}-u\,\rho^2\right) \Delta \breve{a}_x 
 + 2\ell^2\{\breve{a},\Delta \breve{a}\}.
\end{eqnarray}
With $u=B_0^2/\beta \rho^2$, the equation (\ref{modifiedEq}) has exactly the
form of the widely used Q-G equation. 
However, the parameters are essentially different:
Instead of the Rossby radius of deformation $c_g/f_0$, we have the radius
$\rho$ defined in (\ref{Omega}); 
instead of the usual $\beta$-parameter, we have
$B_0^2/\beta \rho^4$.
A similar equation also appears in plasma physics (e.g. 
\cite{Di}),
but again, the physical content and parameters are different. 

\subsection{Estimates for the Earth}
\label{Sect: Estim}

Although the obtained equation (\ref{Eq}) can be applied to different situations
(for different rapidly rotating planets and stars), 
it is interesting to estimate the
parameters (\ref{SmallParam}) for the Earth.

The fine stratification structure 
of the outer core is not clearly known at present \cite{AnnRevCompos}. 
We assume a model stratification, when the liquid outer core consists 
of deep heavier fluid
 and on top of it (near the core-mantle boundary) a shell of lighter
 fluid (the ocean of the core, e.g. \cite{Brag07}).
The upper boundary of the layer is rigid, 
and the lower boundary is moving.

To make estimates, we take the shallow layer depth $H_0\sim 50 km$
and its relative density deficiency 
(negative relative density excess) $10^{-4}$, 
so that the gravity acceleration $g_0\approx 10 m s^{-2}$ is reduced
to the value $g\sim 10^{-3} m s^{-2}$.
This gives the gravity wave speed $c_g\sim 7 m/s$.

At latitudes about $45^\circ$, 
the Coriolis parameter has value $f_0\approx 10^{-4} s^{-1}$, and 
$\beta\approx 3\times 10^{-11}m^{-1}s^{-1}$.
We assume the background toroidal magnetic field 
corresponding to the Alfven speed $B_0\approx 0.3 m/s$. 
Then $\ell\approx 100 km$ and $\rho\approx 150 km$.
Since $R_0\approx 3475 km$, the condition (\ref{regime}) can be well
satisfied; for instance, if ${\mathcal L}\sim 400 km$, the small
parameters (\ref{SmallParam}) take the values
\begin{eqnarray*}
\frac{\beta{\mathcal L}}{f_0}\sim 
\frac{\rho^2}{\mathcal L^2}\sim 0.1,\quad 
\frac{\ell^2}{\mathcal L^2}\sim 0.06;
\end{eqnarray*}
the ratio ${\mathcal B}/B_0$ can be small as well, provided the
nonlinearity level is low.

\section{Energy Distribution}
\setcounter{equation}{0}
\label{Sect: Energy}

\subsection{Energy and Enstrophy. Cascades}
\label{Sect:Cascds}

The equation (\ref{Eq}) --- like its modification (\ref{modifiedEq}) --- 
conserves two positive-definite quadratic integrals
\begin{eqnarray*}\label{EnergyEnstr}
	\int [a^2+(\rho\nabla a)^2]\; dxdy,\quad\quad
	\int[(\nabla a)^2+(\rho\Delta a)^2]\; dxdy.
\end{eqnarray*}
In the usual Q-G situation, 
the first of them is energy, and the second is the enstrophy;
but for the dynamics (\ref{Eq}), the energy and enstrophy switch: 
The second one is the energy, 
corresponding to the energy of the ``shallow water'' MHD (\ref{SMHD}).

We will be interested in cascades in the Fourier space, 
and so, consider the equation (\ref{Eq}) in the Fourier
representation. 
To shorten notations, 
for any wave vector ${\bf k}_j \; (j=1,2,3)$ we keep only its label $j$; 
e.g.\ for the Fourier transform 
\begin{eqnarray*}
a_{\bf k}(t)=\frac{1}{(2\pi)^2}\int a(x,y,t) \; e^{i(p x+q y)}\;dx\,dy
\quad [{\bf k} =(p,q)]
\end{eqnarray*}
we have $a_{{\bf k}_1}\equiv a_1$.
Also, $-j$ stands for $-{\bf k}_j$. 
In the Fourier representation, 
the evolution of a Fourier harmonic with some wave vector, 
say ${\bf k}_1$, is determined by the interaction 
with all other Fourier harmonics ,
and the equation (\ref{Eq}) takes the form
\begin{eqnarray}\label{EqFourier}
\dot a_1&=-\Omega_1 \,a_1 + \int {W}_{-1,2,3}\;a_2\,a_3 \;d{\bf k}_2\,d{\bf k}_3,
\end{eqnarray}
where the dispersion law $\Omega_{\bf k}$ is given in (\ref{Omega}),
and the coupling kernel is
\begin{eqnarray}\label{W123}
\lefteqn{W_{1,2,3}\equiv W({\bf k_1},{\bf k}_2,{\bf k}_3)=}\nonumber\\
 & & \ell^2\, \frac{k_3^2-k_2^2}{1+\rho^2 k_1^2}\;
(p_2 q_3 - p_3 q_2)\;\delta({\bf k}_1+{\bf k}_2+{\bf k}_3).
\end{eqnarray} 
The equation (\ref{EqFourier}) conserves the energy and enstrophy
\begin{eqnarray*}
	\int k^2(1+\rho^2 k^2)\; |a_{\bf k}|^2 \; d{\bf k}, \quad
	\int (1+\rho^2 k^2)\; |a_{\bf k}|^2\; d{\bf k};
\end{eqnarray*} 
if $E_{\bf k}$ is the energy spectrum, 
then $F_{\bf k}=E_{\bf k}/k^2$ is the enstrophy spectrum. 
This implies the direct cascade of energy 
and the inverse cascade of enstrophy 
(contrary to the usual Q-G case).

\subsection{Kolmogorov-type Spectrum for the Inverse Cascade}
\label{Sect:Kolmgrv}

Let us use dimensional considerations 
to find the turbulence spectrum 
for the inverse cascade in the dynamics (\ref{Eq}). 
Dimensional considerations alone are insufficient, 
since the equation (\ref{Eq}) has dimensional parameters. 
However, we can supplement dimensional considerations 
by other arguments in the following two ways.

\subsubsection{Rescaling the dynamical equation}

First of all, let us note that the term with $\rho$ in the equation (\ref{Eq}) 
is small in the considered regime (\ref{regime}), 
and so, can be neglected for dimensional considerations. 
Now, we re-scale the dependent variable $\tilde a =\ell^2 a$,
so that the equation (\ref{Eq}) is reduced to the equation 
with only one dimensional parameter
\begin{eqnarray*}\label{EqRenorma}
\tilde a_t= \frac{B_0^2}{\beta}\Delta \tilde a_x
+2\,\{\tilde a,\Delta \tilde a\};
\end{eqnarray*}
herewith the dimensions of space and time are unaffected.
The corresponding energy spectrum $\tilde E_k = \ell^4 E_k$,
and the corresponding enstrophy flux $\tilde Q=\ell^4 Q$.
Now we assume
\begin{eqnarray*}
\tilde E_k=\left(\frac{B_0^2}{\beta}\right)^\lambda \;
{\tilde Q}^\mu \; k^\nu
\end{eqnarray*}
with undetermined exponents $\lambda,\mu,\nu$.
The turbulence spectrum evolves due to
the third order cumulant 
--- which in weakly nonlinear situations ---
has magnitude proportional to the product of two turbulence spectra.
Therefore, $\tilde Q\propto \tilde E^2$, and $\mu=1/2$. 
We have the following dimensions: \\
{\footnotesize \begin{tabular}{lcr}
the energy spectrum $E_k$  & --- & $m^3/s^2$,\\
the enstrophy spectrum $F_k$ &---& $m^5/s^2$,\\
the energy flux $P$ &---& $m^2/s^3$,\\
the enstrophy flux $Q$ &---& $m^4/s^3$.
\end{tabular}}\\
So, the dimensions of $\tilde E_k$ and $\tilde Q$ are 
$m^7/s^2$ and $m^8/s^3$ respectively. 
Now using dimensional considerations, 
we determine $\lambda=1/2,\,\nu=-3/2\;$ and, returning to the
original variables, find the Kolmogorov-type spectrum (\ref{Ek}).
Interestingly, the spectrum  (\ref{Ek}) is independent of $B_0$.

\subsubsection{Using the scaling of the wave kinetic equation}

We will arrive at the same spectrum (\ref{Ek}) 
when dimensional arguments are supplemented 
by the scaling implied by the wave kinetic equation 
(similar to the dimensional estimates made in \cite{ZakhKS}).

Using the well studied \cite{Kenyon,LoHGill,Reznik,MoPi} 
kinetic equation for the Q-G turbulence (of Rossby waves)  
we have 
(switching the energy and enstrophy spectra)
the wave kinetic equation for our equation (\ref{Eq})
\begin{eqnarray*}\label{KinEq}
\frac{\partial F_1}{\partial t} =
\int \frac{W_{123}(W_{123} F_2 F_3 + W_{231} F_3 F_1 + W_{312} F_1 F_2 ) }
{(1+\rho^2 k_1^2)\,(1+\rho^2 k_2^2)\,(1+\rho^2 k_3^2)}\nonumber\\
\times\delta({\bf k}_1+{\bf k}_2+{\bf k}_3)\;
\delta(\Omega_1+\Omega_2+\Omega_3)\;
d{\bf k}_2\, d{\bf k}_3.
\end{eqnarray*}
We do not care here about numerical factors 
(like $2\pi$ in front of the integral), 
since we will only use this equation for dimensional estimates. 
As well, we can neglect $\rho^2 k^2\ll 1$ 
[due to the condition (\ref{regime})]
in the denominator of the integrand and in $\Omega_{\bf k}$.
The inverse cascade corresponds to the enstrophy flux $Q$; 
the quantity $-Q$ is the enstrophy flux out 
of large sphere $|{\bf k}|\le K$ of big radius $K\rightarrow\infty$, 
and so, $-Q$ is the integral of  $\partial F/\partial t$ over this 
sphere. We also assume that the turbulence
(corresponding to the spectrum that will be obtained) is {\it local}, 
i.e.\ the integral in the kinetic equation converges on this spectrum.
Then the dimensional considerations suggest
\begin{eqnarray*}
Q\sim W^2\; F^2\; k^{-2}\;
\left(\frac{B_0^2}{\beta}k^3\right)^{-1} \!(d{\bf k})^3
  \sim\frac{\ell^4\,\beta}{B_0^2}\, F^2\,k^9,\quad
\end{eqnarray*}
from where we find 
\begin{eqnarray*}
F_{\bf k}\sim\,(\beta\,Q)^{1/2}\,k^{-9/2}
\end{eqnarray*}
and the energy spectrum (\ref{Ek}).

\bigskip

We see from the dimensional considerations 
that the energy integral diverges on the Kolmogorov-type spectrum at small $k$.
This indicates \cite{ZakhKS} that 
the large scales contain most of the energy,
which suggests the following picture. 
Most of the energy follows direct cascade and dissipates at large $k$; 
but a little fraction of the supplied energy 
has to transfer to small $k$ (since enstrophy takes energy);
and the energy piles up there. 
A similar situation holds for the sea wave turbulence, 
where there are direct cascade of energy and inverse cascade of wave action, 
but the large scales contain most of the energy. 

\subsection{Extra Invariant}
\label{Sect:Inv}
A dispersion law $\omega_{\bf k}$ is said to be {\it degenerative}
\cite{ZSch0} if there exists a function $\phi_{\bf k}$ 
such that the equations of resonance triad interactions
\begin{eqnarray}\label{TriadResn}
{\bf k}_1+{\bf k}_2+{\bf k}_3 =0,\quad \omega_1+\omega_2+\omega_3=0 
\end{eqnarray}
imply the equation
\begin{eqnarray}\label{Degene}
\phi_1+\phi_2+\phi_3=0 
\end{eqnarray}
[recall the notational conventions: 
Index $j=1,2,3$ stands for the wave vector ${\bf k}_j$].
The function  $\phi_{\bf k}$ is supposed to be linear independent of
${\bf k}$ and $\omega_{\bf k}$, so that the equation (\ref{Degene})
is not a mere linear combination of equations (\ref{TriadResn}).

The Rossby wave dispersion law
\begin{eqnarray}\label{Rossby}
\omega({\bf k})=\frac{p}{1+\rho^2 k^2}
\end{eqnarray}
turned out to be degenerative \cite{BNZ,B1991}. 
The dispersion law (\ref{Omega}) is a linear combination of functions
(\ref{Rossby}) and $p$, and so, the system (\ref{TriadResn}) --- with
dispersion law (\ref{Rossby}) --- is equivalent to the system 
\begin{eqnarray*}
{\bf k}_1+{\bf k}_2+{\bf k}_3 =0,\quad \Omega_1+\Omega_2+\Omega_3=0
\end{eqnarray*}
with $\Omega_{\bf k}$ given in (\ref{Omega}). 
Therefore, the dispersion law (\ref{Omega}) is also degenerative with
the same $\phi_{\bf k}$, which is equal to the function (\ref{eta}).
Since the function  (\ref{eta}) is continuous (for all ${\bf k}\neq 0$), 
the weakly nonlinear dynamics (\ref{Eq}) possesses 
the extra invariant (\ref{TheIntegral}),
in addition to the energy and the enstrophy. 
The conservation of the extra invariant 
(\ref{TheIntegral}) is {\it adiabatic-like}: 
The quantity $I$ is conserved approximately over long (nonlinear) time. 
(The extra invariant and its significance for
geophysical fluid dynamics are reviewed in \cite{balk13}).

\subsection{Inverse Cascade}
\label{Sect:InvrsCascd}

We will now show that for the dynamics (\ref{Eq}),
the presence of the extra invariant implies 
the energy accumulation 
in the sector $60^\circ <|\theta|< 90^\circ$
(where $\theta$ is the polar angle of the wave vector ${\bf k}$).
\footnote{Since $a_{\bf k}=a_{-{\bf k}}^\ast$, 
for considerations of the turbulence spectrum, 
the set $60^\circ<|\theta|<90^\circ$ is equivalent 
to the sector $60^\circ<\theta<120^\circ$, and we call the former set
also a {\it sector}.}

A similar fact was derived \cite{B2005}
for the turbulence of Rossby waves, but the reasoning here is
different. This is because the energy cascade has switched direction.

\begin{figure}
\includegraphics{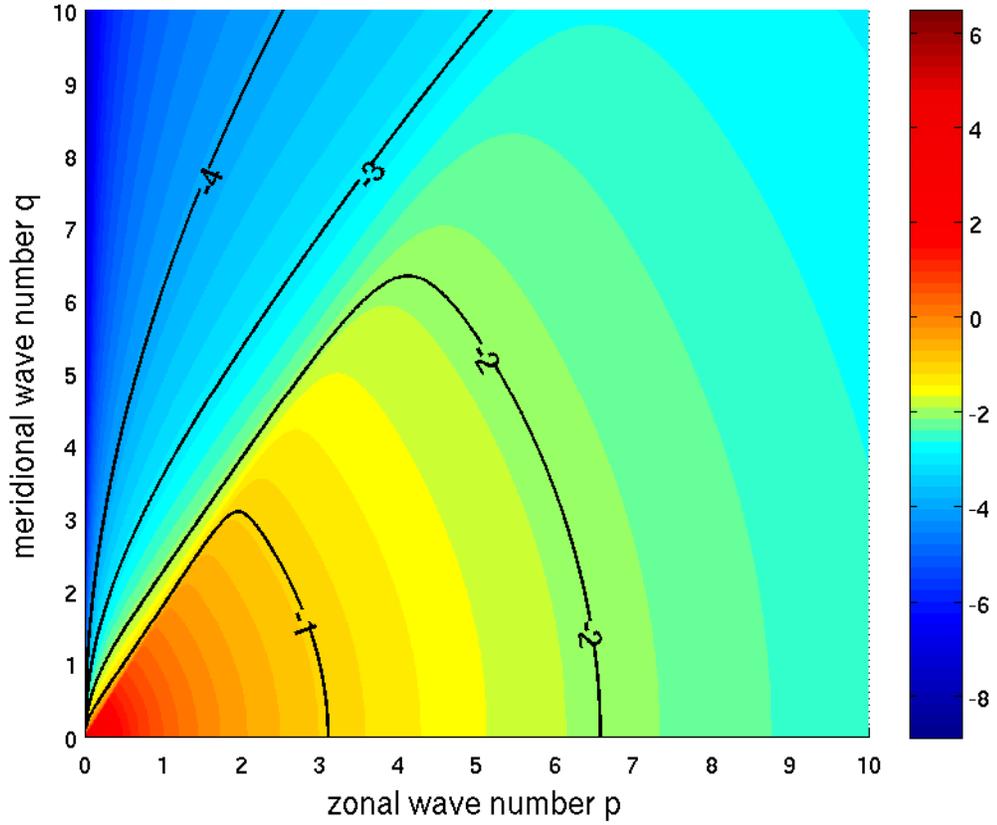}
\caption{The values of $\log_{10}(\varphi)$, see (\ref{J}).
The black curves are the level lines $\varphi({\bf k})=10^n$
for 4 integers $n=-1,-2,-3,-4$. 
All contour lines pass through the origin tangent to the
$q$-axis. For this figure, the parameter $\rho=0.01$ 
[so that indeed $\rho^2k^2\ll 1$, 
even for the mode with the biggest $k$, when ${\bf k}=(10,10)$] and
$\beta/B_0^2=1$.}
\label{fig:BalArg}
\end{figure}

When $p\rightarrow 0$, 
the function (\ref{eta}) behaves like $\Omega_{\bf k}/k^2$ 
(up to a constant factor), and so, we linearly combine $I$ and $F$ to
eliminate the common asymptotics
\begin{eqnarray}\label{J}
J=I-\frac{2\sqrt{3}\rho\beta}{B_0^2} F =
 \int \varphi({\bf k})\; E_{\bf k}\; d{\bf k},\quad\mbox{where}\nonumber\\
\varphi({\bf k})=\frac{\eta({\bf k})}{\Omega_{\bf k}} -
\frac{2\sqrt{3}\rho\beta}{B_0^2}\frac{1}{k^2}\;\;
\left[=\; O(p^2),\; p\rightarrow 0\right].\quad
\end{eqnarray}
Obviously, the integral $J$ is also an invariant of dynamics (\ref{Eq}), 
just like the integral $I$; from now on, we will deal with $J$ instead of $I$.
Figure \ref{fig:BalArg} shows the function $\varphi({\bf k})$. 

Asymptotically for $k\rightarrow 0$ (up to a constant factor $\beta/B_0^2$), 
\begin{eqnarray}\label{asymp}
\varphi({\bf k})\sim\left\{\begin{array}{lc}
\mbox{\Large$\frac{\pi}{p k^2}$},&\;\; |\theta|<60^\circ\,,\\
\mbox{\Large$\frac{8\sqrt{3}\, \rho\; p^2}{(q^2-3p^2) k^2}$},&\;
60^\circ<|\theta|<90^\circ\; ;
\end{array}\right.
\end{eqnarray}
so, $\varphi\sim k^{-3}$ in the sector $|\theta|<60^\circ$,
and $\varphi\sim\rho\, k^{-2}$ in the sector $60^\circ<|\theta|<90^\circ$.
The expression (\ref{asymp}) gives a simple approximation 
for the function (\ref{J}), away from the lines $q=\pm\sqrt{3}\,p$.

Let us consider a simple model situation, 
when the energy is generated at some scale $k_0$ at a rate ${\mathcal E}_0$
and dissipated at some scales $k_1$ and $k_2$ 
at rates ${\mathcal E}_1$ and ${\mathcal E}_2$ respectively. 
We assume $k_1\ll k_0\ll k_2$, 
so that both direct and inverse cascades are realized.
The conservation of the energy and the enstrophy gives 
\begin{eqnarray*}
{\mathcal E}_0={\mathcal E}_1+{\mathcal E}_2, \quad
\frac{1}{k_0^2}{\mathcal E}_0=
\frac{1}{k_1^2}{\mathcal E}_1+\frac{1}{k_2^2}{\mathcal E}_2,
\end{eqnarray*}
from where we have 
\begin{eqnarray}\label{E1}
{\mathcal E}_1=\frac{1/k_0^2-1/k_2^2}{1/k_1^2-1/k_2^2}{\mathcal E}_0
\approx\left(\frac{k_1}{k_0}\right)^2 {\mathcal E}_0\ll {\mathcal E}_0,
\end{eqnarray}
so that indeed a small fraction of the generated energy is transferred
towards the origin (most of the energy follows the direct cascade
towards large $k$).

We will now arrive at a contradiction if the energy ${\mathcal E}_1$ 
ends up in the sector $60^\circ<|\theta|<90^\circ$. 
Indeed, the dissipation of energy ${\mathcal E}_1$ in this sector 
is accompanied by dissipation of the extra invariant $k_1^{-3}{\mathcal E}_1$; 
the latter cannot exceed the total amount of the extra invariant 
generated at scale $k_0$: $k_0^{-3}{\mathcal E}_0$ 
(if some energy was generated in the sector $60^\circ<|\theta|<90^\circ$, 
then the generated amount of the extra invariant is even smaller). 
Thus, $k_1^{-3}{\mathcal E}_1<k_0^{-3}{\mathcal E}_0$, 
which upon substitution of 
${\mathcal E}_1$ in terms of ${\mathcal E}_0$ from (\ref{E1}) 
leads to contradiction $k_1>k_0$. 
The contradiction still remains 
if only a significant part (not the entire amount) of ${\mathcal E}_1$ 
ends up in the sector  $60^\circ<|\theta|<90^\circ$, 
but a longer inertial interval might be required.

\subsection{Zonal Flow}
\label{Sect:Zonl}

We will now see that for some energy input at the scale $k_0$, the
region of energy accumulation is more narrow: The energy accumulates in
the vicinity of the $q$-axis, which corresponds to zonal flow
(alternating zonal jets).

\begin{figure}
\vspace*{5mm}
\begin{center}
\includegraphics{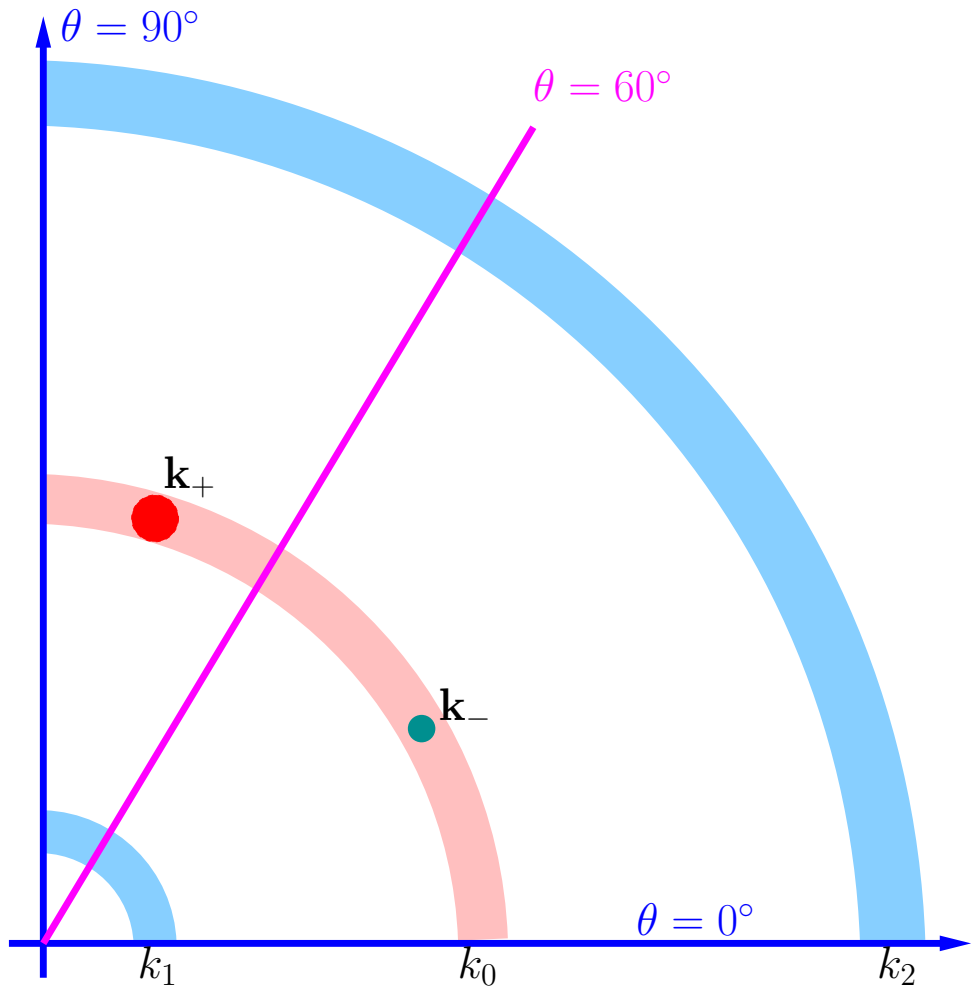}
\end{center}
\caption{The energy input at the scale $k_0$ consists of two parts: 
The energy is generated in the sector $60^\circ<|\theta|<90^\circ$,  
but dissipated in the sector $|\theta|<60^\circ$ ($\theta$ is
the polar angle of the wave vector ${\bf k}$). 
}
\label{fig:squeeze}
\end{figure}

Suppose that the energy input at the scale $k_0$ consists of 
the energy source in the sector $60^\circ<|\theta|<90^\circ$,
and the energy sink in the sector $|\theta|<60^\circ$ 
(see Fig.\ \ref{fig:squeeze}), 
in such a way that the input of the extra invariant is small, 
while the energy input is significant. Such input is possible
because the extra invariant is anisotropic.

To illustrate this situation, let us 
continue with the above model situation, 
assuming that the energy is generated at some ${\bf k}_+$ 
at a rate ${\mathcal E}_+$ (in the sector $60^\circ<|\theta|<90^\circ$),
but dissipated at some ${\bf k}_-$
at a rate ${\mathcal E}_-$ (in the sector $|\theta|<60^\circ$); 
$|{\bf k}_+|\sim|{\bf k}_-|\sim k_0$; the dissipation at scales
$k_1$ and $k_2$ is still present; Fig.\ \ref{fig:squeeze}.
The energy input is ${\mathcal E}_0={\mathcal E}_+-{\mathcal E}_-$,
and the extra invariant input is 
$\varphi_+{\mathcal E}_+-\varphi_-{\mathcal E}_-$. 
The function $\varphi({\bf k})$ has much bigger values in the
sector $|\theta|<60^\circ$ than in the sector
$60^\circ<|\theta|<90^\circ$. 
So, if ${\mathcal E}_+\gg{\mathcal E}_-$,
the energy input will be significant,
while the extra invariant input can be small:
\begin{eqnarray*}
\frac{\varphi_+{\mathcal E}_+-\varphi_-{\mathcal E}_-}
     {\varphi_+{\mathcal E}_++\varphi_-{\mathcal E}_-}
\ll 1.
\end{eqnarray*}
When the energy (and the enstrophy) dissipate at the scale $k_1$,
only a small amount of the extra invariant $J$ can be dissipated (less
than the total supplied amount of $J$ at scale $k_0$). This is only possible if
the dissipation occurs near the $q$-axis 
[where the function $\varphi({\bf k})$ has small magnitude; 
$\varphi({\bf k})=O(p^2)$ as $p\rightarrow 0$, see (\ref{J})]. Thus, the
accumulated energy must squeeze near the $q$-axis, which corresponds
to zonal flow.

It is interesting to consider this zonal flow generation in light of
the results on the $\beta$-plane MHD \cite{TobiasDiamond}. 
To investigate the Solar tachocline, 
they considered two-dimensional MHD on the $\beta$-plane 
and looked for the generation of zonal flow. 
Such flow could limit the angular momentum transport 
in the solar tachocline and would partially explain
the tachocline confinement. 
However, using direct numerical simulations, 
they came to the conclusion that 
``in the absence of magnetic fields, 
nonlinear interactions of Rossby waves lead to 
the formation of strong mean zonal flows; 
but {\it the addition of even 
a very weak toroidal field suppresses 
the generation of mean flows}.'' 

A similar conclusion was reached by 
for the MHD on a rotating spherical surface \cite{TDMarston}. 
We should note that 
there is no contradiction between these results and the present paper: 
They consider the opposite limiting situation, 
when ${\mathcal L}\ll\rho$ --- cf.\ (\ref{regime}).

\subsection{Remark on More General Dispersion Law}
 
Some other scaling of the ratios (\ref{SmallParam}) 
[different from the one taken in the present paper] 
can bring into consideration the terms with time
derivatives in (\ref{QGv}). 
Then the dispersion law would be more general
\begin{eqnarray}\label{GenDisp}
\Omega_{\bf k}=\frac{B_0^2}{\beta} \frac{p k^2}{1+\rho^2k^2+\ell^4k^4}.
\end{eqnarray}
It is interesting that in both limits of long and short waves, this
dispersion law is degenerative and admits an extra invariant. Indeed,
\begin{eqnarray*}
\Omega^{\mbox{\footnotesize long}}_{\bf k}=\frac{B_0^2}{\beta} 
                       \frac{p k^2}{1+\rho^2k^2},\quad
\Omega^{\mbox{\footnotesize short}}_{\bf k}=\frac{B_0^2}{\beta} 
                       \frac{p}{\rho^2+\ell^4k^2}
\end{eqnarray*}
are both Rossby dispersion laws (Doppler shifted in the first case).
However, it seems unlikely that the general dispersion law (\ref{GenDisp})
is degenerative. Nevertheless, it appears possible to establish extra
conservation in the generalized situation using the smallness of the
parameter $\ell$ (so that the approximate conservation of the extra
invariant remains within the same bound).

\section{Conclusion}

We have derived a single equation [Eq (\ref{Eq})]
for the {\it slow large-scale} MHD 
of a shallow fluid layer on the beta-plane 
of a rapidly rotating planet or star. 
This equation (after Doppler shift) 
has the form of the quasi-geostrophic equation 
(familiar in Geophysical fluid dynamics) 
or the Hasegawa-Mima equation (familiar in Plasma physics), 
but the physical content and 
the parameters are very different --- Sect. \ref{Sect: Eq}. 
The validity regime of this equation fits 
{\it Earth's ocean of the core} --- 
a layer of lighter fluid at the core-mantle boundary 
--- Sect. \ref{Sect: Estim}.
 
The derived equation implies two cascades: 
the direct energy cascade and the inverse enstrophy cascade 
(contrary to the situation in the usual Q-G equation) 
--- Sect. \ref{Sect:Cascds}.
Using the dimensional considerations, we have found the
Kolmogorov-type spectrum for the inverse cascade. 
This spectrum indicates that larger scales contain most of the
energy (Sect. \ref{Sect:Kolmgrv}) in agreement 
with experimental observations;
this provides evidence that the obtained
equation  (\ref{Eq}) is physically plausible.

The equation (\ref{Eq}) possesses an extra invariant 
(in addition to the energy and enstrophy) 
--- Sect. \ref{Sect:Inv}. 
Its presence implies the energy accumulation 
in the sector $60^\circ<|\theta|<90^\circ$
--- Sect. \ref{Sect:InvrsCascd}. 
The extra invariant can also imply
the formation of zonal flow if the energy input includes 
the energy source in the sector $60^\circ<|\theta|<90^\circ$  
and its sink in the sector $|\theta|<60^\circ$ --- Sect. \ref{Sect:Zonl}. 

Zonal flows also take place in magnetized plasmas; 
they improve the particles and heat confinement in tokamaks, e.g. \cite{Di}.
Unlike Earth's outer core, in plasmas, 
we have some control over the forcing and dissipation. 
The similarity of equation (\ref{Eq}) 
to the Hasegawa-Mima equation 
suggests a way to generate zonal flows in plasmas. 
To set up a strong zonal flow, 
one needs to supplement the energy generation (at scale $k_0$)
by its dissipation in the sector $|\theta|<60^\circ$ 
(see Fig.\ \ref{fig:squeeze}); 
herewith one should supply significant energy amount while supplying
small amount (as little as possible) of the extra invariant.
Such input is feasible,
since the extra invariant is essentially anisotropic.
It is interesting that the inclusion of the dissipation, 
far away from the $q$-axis, can lead to very focused zonal jets, 
when the energy is accumulated very tightly around the $q$-axis
(which corresponds to zonal flow). 
It is unclear how small the input of the extra invariant can be achieved,
since we only control increment/decrement, 
but not the actual generation/dissipation.
I plan numerical simulations of
the Hasegawa-Mima equation and similar equation (\ref{Eq}) 
to investigate how the inclusion of dissipation
into the energy input
affects the formation of zonal flows in plasmas 
and in the large scale quasi-geostrophic MHD.

\bigskip
I wish to thank Peter Weichman, 
Yuan-Pin Lee, and Lance Miller for beneficial discussions.

\bibliographystyle{unsrt}
\bibliography{My}
\end{document}